\documentclass[5p]{elsarticle} 
\usepackage{graphicx,amssymb,amsmath,subfigure} 
\journal{Astroparticle Physics} 
\begin{document} 
\begin{frontmatter} 
\title{Detailed Studies of Atmospheric Calibration in Imaging Cherenkov Astronomy} 
\author[Durham]{Nolan, S.J.}
\ead{s.j.nolan@dur.ac.uk}
\author[Gerd]{P\"uhlhofer, G \fnref{fn1}}
\author[Durham]{Rulten, C.B.}
\address[Durham]{Department of Physics, Durham University, South Road, Durham,  DH1 3LE, United Kingdom} 
\address[Gerd]{Landessternwarte, Universit\"at Heidelberg, K\"onigstuhl, D 69117 Heidelberg, Germany}

\fntext[fn1]{Now at Institute f\"ur Astronomie und Astrophysik, Sand 1, D-72076 T\"ubingen, Germany}
\begin{abstract} 
\noindent The current generation of Imaging Atmospheric Cherenkov telescopes are allowing the sky to be probed with greater sensitivity than ever before in the energy range around and above 100 GeV. To minimise the systematic errors on derived fluxes a full calibration of the atmospheric properties is important given the calorimetric nature of the technique. In this paper we discuss an approach to address this problem by using a ceilometer co-pointed with the H.E.S.S. telescopes and present the results of the application of this method to a set of observational data taken on the active galactic nucleus (AGN) PKS 2155-304 in 2004. 
 \end{abstract} 
\end{frontmatter}
\section{Introduction}
\noindent Astronomy between 100 GeV and a few tens of TeV has been possible for many years now thanks to Imaging Atmospheric Cherenkov telescopes (IACTs). The field  has benefited immensely from the advent of the current generation of telescope systems, such as H.E.S.S. \cite{HESS}, VERITAS \cite{VERITAS}, MAGIC \cite{MAGIC}, and CANGAROO III \cite{CANGAROOIII}. Since 2002, the number of gamma-ray sources known to emit above 100 GeV has increased from a handful to over 100 \cite{review},\cite{review2}. These sources fall into two broad categories:  those which  produce a constant flux of gamma-rays (e.g. supernova remnants, pulsar wind nebulae), and variable sources, which produce either a periodically changing flux of gamma-rays (pulsar wind binary systems) or a highly irregular flux  (active galactic nuclei). Precise flux and energy spectra calculations for these discoveries can suffer from one major potential problem. The fact is that the atmosphere itself is the detector; it is where the particle shower is initiated by the incident gamma-ray and the medium through which the Cherenkov photons must travel. The energy estimation of an individual gamma-ray is based around calorimetry of the shower, which in turn is based on Cherenkov photon deposition, therefore any change in atmospheric quality can affect the signal detected.\newline

\noindent Unfolding the effect of a varying atmosphere from the lightcurve of variable sources is a particularly tricky task, as it is vital to distinguish flux variations at the source from variations induced due to atmospheric changes. Under normal operations, most IACTs utilise the uniform rate of background cosmic-ray initiated showers as a measurement of atmospheric clarity. In addition, some experiments also operate atmospheric monitoring equipment such as infrared radiometers (to measure sky temperature), weather stations, and equipment for measuring the optical properties of the atmosphere such as lidars, transmissometers, and starlight monitors. 

Usually, the information from these instruments is applied alongside the cosmic-ray rate as a selection parameter to identify good data, meaning that the data are just discarded if they fail predefined weather quality cuts. More sophisticated techniques aim to resurrect otherwise unusable data. Such methods involve the scaling of the gamma-ray image brightness before application of the standard energy reconstruction and flux determination scheme \cite{Dorner}, or scaling of the gamma-ray rate itself to obtain corrected flux values \cite{Jamie}. These relative corrections have been applied successfully in the literature, but both methods have their specific shortcomings. Rate scaling does not restore the original energy scale and strictly speaking only works for power-law gamma-ray spectra (of known photon index). Image brightness scaling restores the original event energy but the loss of non-triggered events near detector threshold needs to be compensated. The latter issue is particularly tricky for Cherenkov telescopes, since events close to \textbf{detector} threshold at large distance from the telescope contribute a sizable fraction to spectral bins well above the \textbf{energy} threshold of the instrument.

Ultimately, the entire energy and flux reconstruction scheme of an instrument can be recalibrated using atmospheric transmission tables that are adapted to the atmospheric conditions during data taking. This approach is pursued for the first time for Cherenkov telescopes in the work presented here. A drawback of the method is that it is very computationally intensive, because for each atmospheric condition a completely new set of simulations and detector calibration needs to be produced. The ultimate advantage of the method is that, in principle, data taken under arbitrary atmospheric conditions could be analyzed, if the atmospheric conditions were continuously monitored with sufficiently high precision. 

The work presented in this paper aims to completely restore an extensive data set taken with the H.E.S.S. telescope array in August and September 2004. The array was observing the Active Galactic Nucleus PKS 2155-304 in the context of a multiwavelength campaign at that time. Already by eye, the observers noted a large population of low-level aerosols during data taking. More quantitatively, we use the data from a commercially manufactured ceilometer located at the H.E.S.S. site to constrain the planetary boundary layer and help modeling the atmospheric transmission between the site and this layer. As ultimate indicator of the atmospheric condition of each observing run, the cosmic-ray induced background trigger rate is invoked, which is a measure of the (relative) energy threshold of the instrument. We use the match of simulated and observed cosmic-ray background trigger rate as figure of merit to see whether the aerosol content used in the simulation is correctly adapted to the data. Once a match is arrived at, the transmission tables are then folded with gamma-ray simulations to produce recalibrated, atmosphere-corrected lookup tables for energy and effective area.

The paper is organized as follows: In section 2, we will discuss the influence of varying atmospheric conditions on the air shower images, and present the ceilometer data that are used to constrain the low-level aerosols that dominate the conditions of the presented H.E.S.S. data set. In section 3, the required atmospheric modeling is briefly introduced. In section 4, we demonstrate that the hypothetic existence of high-level aerosols or clouds in the data can effectively be excluded by looking at the reconstructed shower height of the air shower data themselves. Section 5 describes how the gamma-ray reconstruction is affected by the adapted low-level aerosol contents. In section 6, some key results of the restored PKS 2155-304 data set are discussed; the complete set of results will be presented in a forthcoming H.E.S.S. collaboration paper. In section 7, we compare the results to a small set of similarly affected Crab data, and discuss possible improvements of the technique, before concluding in section 8.

While the analysis presented in this paper is limited to a specific case and does not aim at establishing a universally adaptable restoration scheme for any data affected by arbitrary bad weather, it could present the first step towards such a machinery.

\begin{figure*}
\centering 
\includegraphics[width=14.75cm]{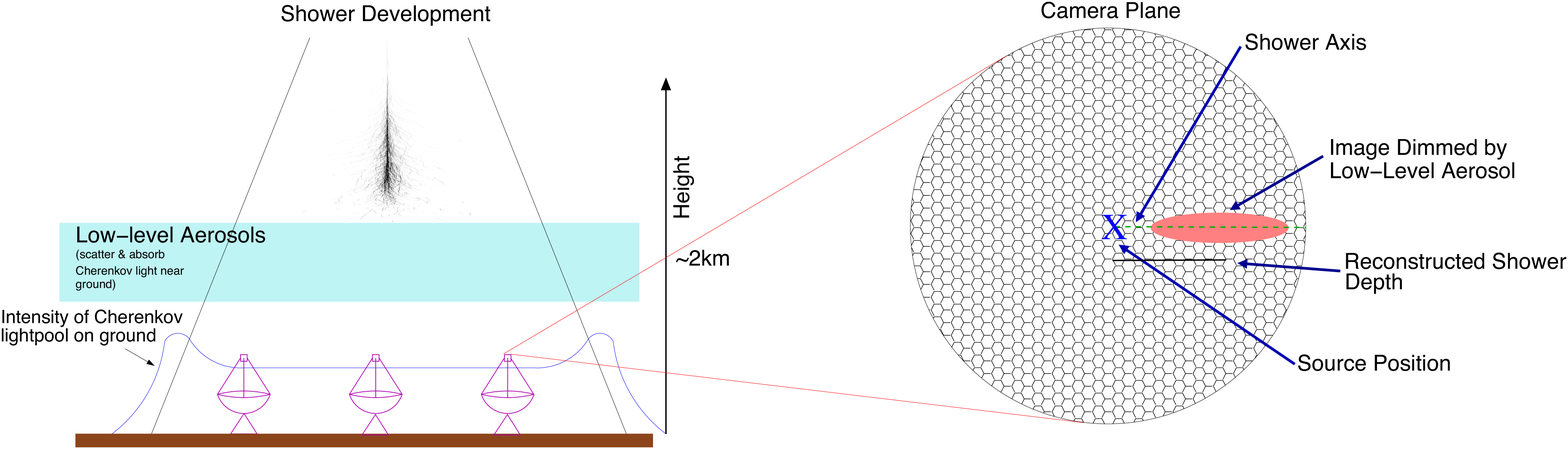}
\includegraphics[width=14.75cm]{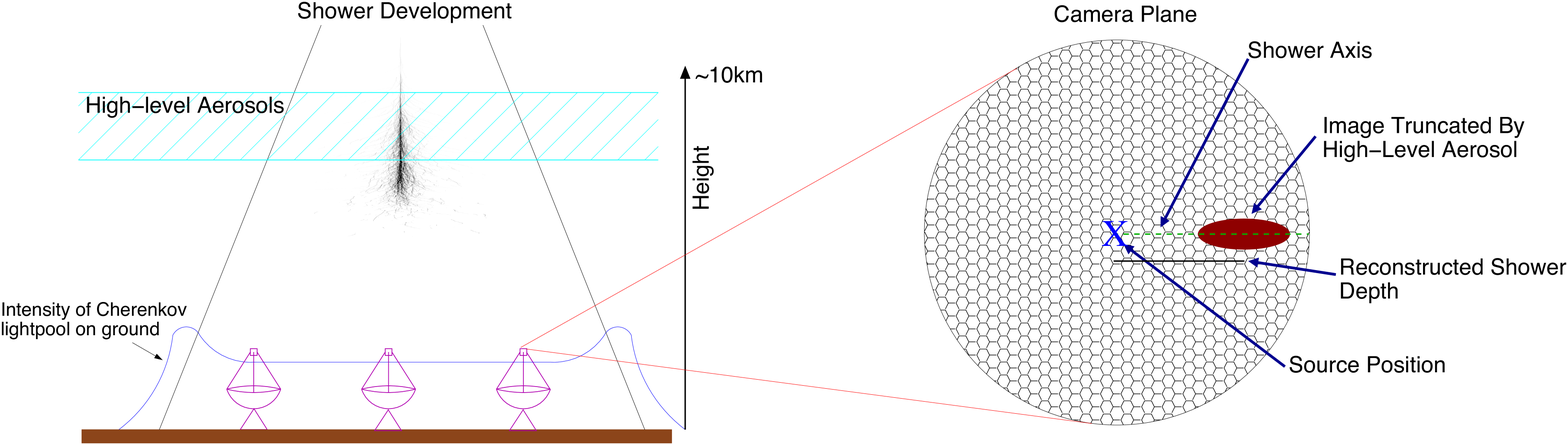}
\caption{{\it Top Panel:} Schematics of gamma-ray shower development showing relative position of low-level aerosol populations. Indicated in addition is the Cherenkov photon density at ground level and the effect on gamma-ray Cherenkov image as seen by telescope camera. {\it Bottom Panel:} Same is shown for high-level aerosol populations.} 
\label{3figs} 
\end{figure*} 

 \section{Atmospheric Monitoring}
\noindent The atmosphere affects the atmospheric conversion of a primary gamma-ray to Cherenkov photons at the telescopes' location in several ways. The (molecular) density profile of the atmosphere determines at which height a shower develops, and is also directly related to the refractive index of air, which controls both the angle under which Cherenkov light is emitted  and the quanitity of Cherenkov photon emission. Last but not least, the transmission of Cherenkov photons from the shower to the telescopes is determined by the density profile of molecules (Rayleigh scattering, with a crucial impact of H2O) and aerosols (Mie scattering).\newline

\subsection{Atmospheric profile}
\noindent Regular simulations of the H.E.S.S. array are based on standard desert atmospheric conditions, as represented by the MODTRAN 4 desert aerosol model, with a default windspeed of 10 metres per second. For the work presented here, it was important to see if relative changes in the molecular density  profile might be expected. This would lead to changes in the cosmic ray rate (again through a change of the energy threshold of the detector), which would in the suggested approach then be misinterpreted as a change of atmospheric transparency. A check on the seasonal stability of the molecular profile of the atmosphere was therefore performed. This was done using British Atmospheric Data Centre  radiosonde data taken from launches conducted at Windhoek airport, located approximately 70 kilometres North East from the H.E.S.S. site \cite{BADC}. By performing cosmic-ray simulations using radiosonde data for the Namibian summer and winter, a difference in simulated cosmic-ray trigger rate for the H.E.S.S. telescope system of  $<5\%$ was seen, due to the changing density (and hence refractive index) profile of the atmosphere. Given the relative stability of the molecular profile for Namibia within a given season, its effect on the work presented in this paper is deemed negligible.

\subsection{Scattering of Light}
\noindent A population of scattering particles can affect the data in two ways, as shown schematically in Figure \ref{3figs}. Aerosols near ground-level lower the yield of Cherenkov light from a shower of given energy and impact parameter, thus making the telescopes trigger less frequently and the camera images dimmer. The image shape is however not affected. As an array of IACTs such as H.E.S.S. performs a trigonometric reconstruction of the impact parameter of the primary gamma-ray, the event will still be reconstructed as coming from the source direction, but will (if uncorrected) be systematically misidentified with a lower reconstructed energy than its true energy (see section 5).\newline
In the context of the presented analysis, it was important to verify that all aerosols are below the Cherenkov emission height, to make sure that cosmic-ray and gamma-ray showers (having different penetration depths) are similarly affected. For this purpose, we exploited the fact that if a significant aerosol population occurs at (or near) shower maximum, then not only the brightness but also the shape of the camera images will be affected. This is then seen mainly in the reconstructed shower height, for which the result of a simulated data-set will be presented in Section 4. \newline
We note in passing that, as has been noted by others \cite{KonradAtm}, for a limited field of view instrument such as a Cherenkov telescope, the importance of multiple scattering is negligble, given that any scattered light will be lost from the field of view. 

\subsection{Ceilometer Measurements}
\noindent The H.E.S.S. collaboration operates several lidar devices at their Namibian site. In this paper we will focus on data derived from the commercially built Vaisala CT25K ceilometer, which was the first to be installed at the telescope site; its specifications are given in Table \ref{tab1} and detailed documentation can be found in \cite{ct25k}.\newline

\begin{table}[h]
\begin{center}
{\footnotesize
\begin{tabular}{| c  |  c |}
\hline
Parameter&Value\\
\hline
Wavelength&905 nm\\
Frequency&5.57 kHz\\
Pulse Width&100 ns\\
Energy/Pulse& 1.6$\mu$J\\
Range&50 m to 7.5 km\\
Spatial Resolution&16 m\\
Laser Class&4\\
\hline
\end{tabular}
}
\end{center}
\caption{\label{tab1}Specifications of the CT25K ceilometer \cite{ct25k}.}   
\end{table}

\noindent Ceilometers are somewhat basic lidar instruments that are often used to report cloud layers, vertical visibility and can also be used to constrain the planetary boundary-layer \cite{pbl}. The CT25K device was operated at the H.E.S.S. site from 2002 to 2007 and atmospheric measurements were taken co-pointed with the H.E.S.S. telescopes for a large fraction of this time. Table \ref{tab2} below provides a summary of the ceilometer data recorded simultaneously with observations of PKS 2155-304. Overall, approximately $\backsim 73\%$ of the PKS 2155-304 observation runs used in this paper have simultaneous ceilometer data.\newline

\begin{table}[h]
\begin{center}
{\footnotesize
\begin{tabular}{| c | c | c | c | c |}
\hline
Atm Model & Good Runs & Bad Runs & Total Runs & Bad $\%$ \\ [0.5ex]
\hline
17.5 & 45 & 3 & 48 & $\backsim 6\%$ \\[0.5ex]
20.0 & 21 & 25 & 46 & $\backsim 54\%$ \\[0.5ex]
22.5 & 53 & 15 & 68 & $\backsim 22\%$ \\[0.5ex]
All & 119 & 46 & 162 & $\backsim 27\%$ \\[0.5ex]
\hline
\end{tabular}
\caption{\label{tab2} Summary of ceilometer data recorded simultaneously with observations of PKS 2155-304. Each unit represents a single 28 minute observation run. Runs with ceilometer data are denoted 'Good' and without ceilometer data 'Bad'. Overall, approximately $73\%$ of the PKS 2155-304 observation runs have simultaneous ceilometer data.}
}
\end{center}
\end{table}
\noindent The CT25K ceilometer proved to be a particularly stable device that is easily portable making it an ideal instrument for atmospheric quality site-testing of future ground-based gamma-ray telescope arrays \cite{cta}. However the ceilometer was primarily designed to study cloud height and its ability to calculate accurate extinction values for low-level aerosol populations is limited. Furthermore, the instrument's laser wavelength and range were sub-optimal for Cherenkov astronomy, but in everyday use the system was ideal for spotting incoming clouds, or as in the case of the methods discussed later, for locating low-level aerosol populations and constraining the planetary boundary layer.\newline

\begin{figure}[tbh]
\begin{center}
\includegraphics[clip=,angle=0,width=0.48\textwidth,height=0.30\textheight]{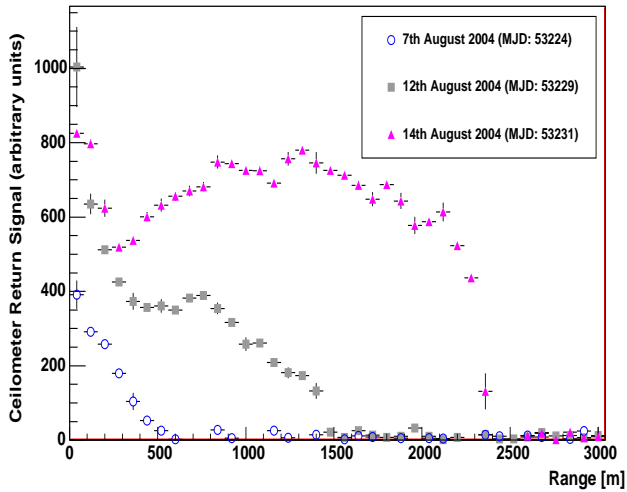}
\caption{The derived return signal as seen by the CT25K ceilometer for several nights in August 2004. The 7th August shows the normal clear-sky behaviour, whereas the other nights show increasing low-level aerosol density. The return signal here is a derived quantity (proportional to extinction), produced by closed source software. }
\label{fig1}
\end{center}
\end{figure}

\noindent Although the H.E.S.S. site in Namibia is largely unaffected by severe weather and benefits from very good astronomical seeing, each year during August and September a significant dust layer (thought either to be due to sand blown up from the nearby Namib desert, or local bush fires) is occasionally seen over the site. The ceilometer clearly identifies such events as shown in Figure \ref{fig1}, where the increased return signal is inferred to originate from an increase in aerosol density. The density of the aerosol layer within the data-sets discussed herein was seen to be stable during a given night, but exhibits night by night variability.\newline

\noindent These layers greatly affect the array trigger rate for cosmic rays which drops by around 70$\%$ at worst. Once the zenith angle dependence of the cosmic-ray trigger rate (due to increasing energy threshold with zenith angle) has been removed, the trigger rate variation for the data-set discussed in a given night is relatively small (typically $< 10\%$), again indicating little change in atmospheric quality during a night. However significant changes are observed on a night to night basis. A possible reason for this overall behaviour could be that changes in low-level aerosol density are driven by diurnally changing wind patterns.\newline

\begin{figure}[tbh]
\begin{center}
\includegraphics[width=0.48\textwidth,height=0.3\textheight]{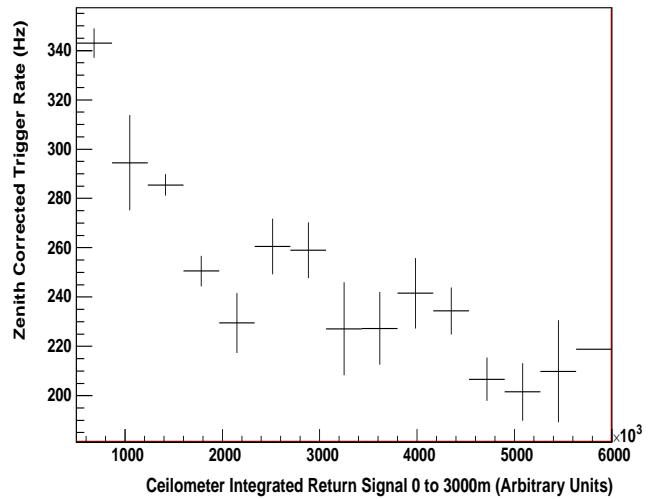}
\caption{The correlation between array cosmic-ray trigger rates recorded at the H.E.S.S. site and integrated return signal recorded with the ceilometer. This plot shows that as the low-level aerosol density increases so the array trigger rate decreases.}
\label{ceilo_corr}
\end{center}
\end{figure}

\noindent Figure \ref{ceilo_corr} illustrates the correlation between array cosmic-ray trigger rates recorded at the H.E.S.S. site and integrated return signal measurements recorded with the ceilometer. This plot shows that as the low-level aerosol density increases so the array trigger rate decreases.

\section{Atmospheric Modelling}

\begin{figure*}[t]
\begin{center}
\includegraphics[clip=,angle=0,width=0.90\textwidth,height=0.30\textheight]{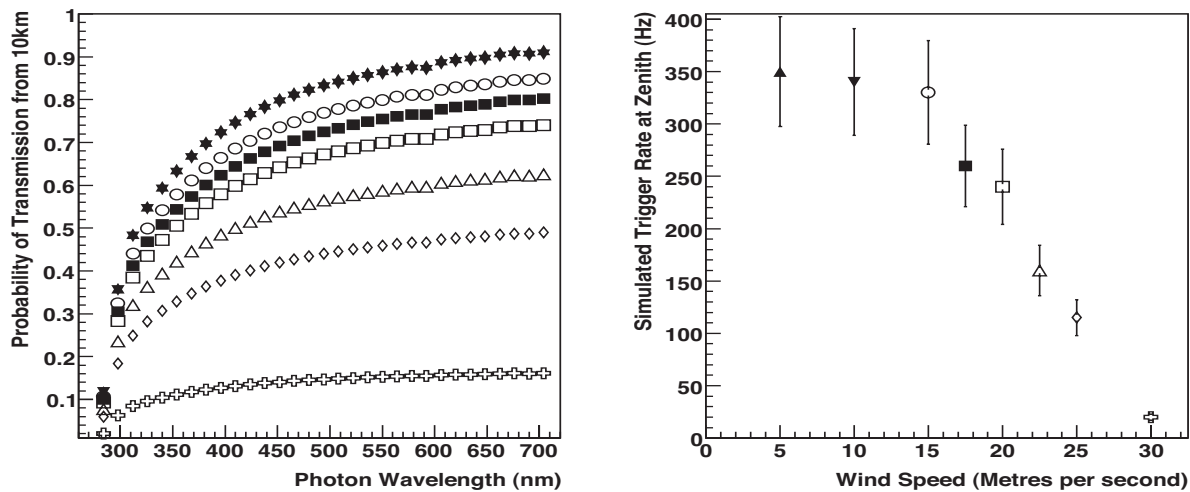}
\caption{{\it Left Hand Panel:} The transmission of light from an altitude of 10 km versus wavelength for different MODTRAN4 wind speeds ranging from 5 m/s (filled upward triangle) to 30 m/s (open cross) (in steps of 5 m/s). {\it Right Hand Panel:} The relationship between the H.E.S.S. simulated cosmic-ray trigger rate at zenith and increasing low-level dust (set using the wind speed parameter). The errors are purely driven by systematic effects due to uncertainties in the folded cosmic-ray flux at 1 TeV taken from \cite{Wiebel}.   }
\label{wssfig}
\end{center}
\end{figure*}

\noindent During the dusty episodes described in Section 2 the position of the low-level aerosol layer could be extracted from the CT25K ceilometer data. The MODTRAN4 simulation code has been used to produce  all models of vertical aerosol structure (plus the molecular absorption and scattering) cited within this paper \cite{Modtran}. The $desert$ aerosol model within MODTRAN4 introduces a layer of aerosols (sand particles) of depth 2 km directly above ground level, the density of which is increased as the wind speed parameter is increased. For the work presented here, \textbf{the wind speed is a tuning parameter} that enables the simultaneous matching of cosmic-ray trigger rate and image parameter distributions, and is not a reflection of the measured wind speed at the site. The effect of changing the wind speed parameter on the probability of transmission of a photon from an altitude of 10 km to the telescope altitude (1.8 km) and the resultant effect on the simulated cosmic-ray rate is shown in Figure \ref{wssfig}, where wind speed is varied from 5 to 30 m/s in steps of 5 m/s.\newline

\begin{figure*}[t]
\begin{center}
\includegraphics[clip=,angle=0,width=0.95\textwidth,height=0.40\textheight]{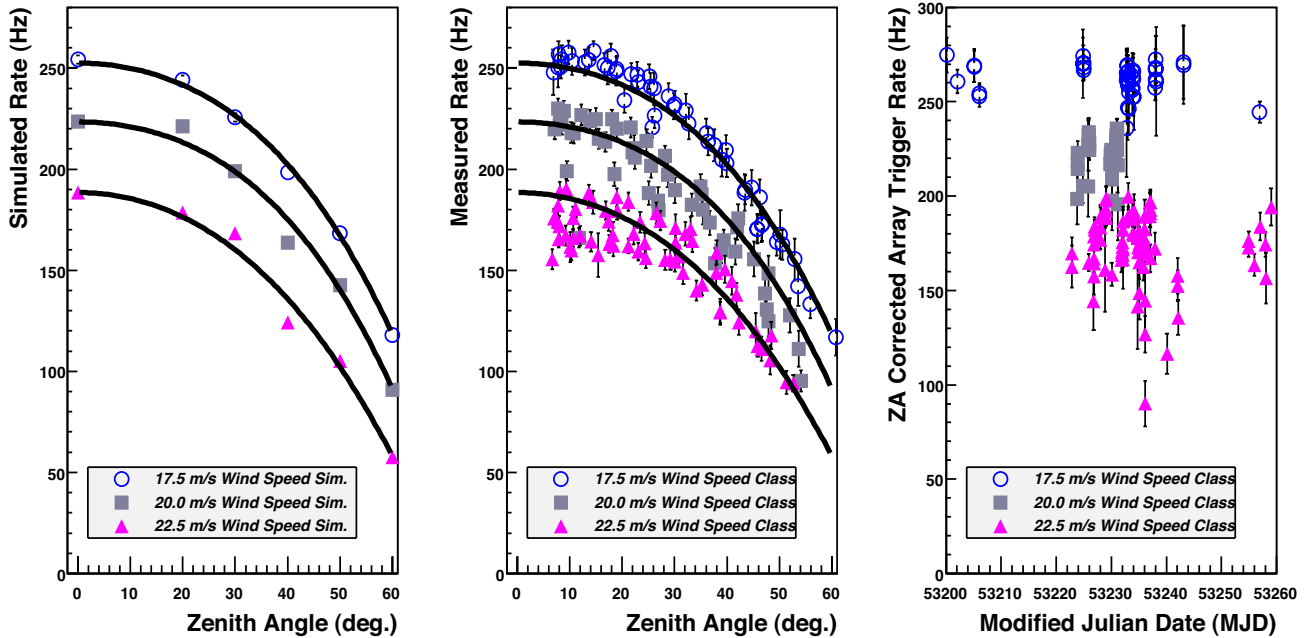}
\caption{{\it Left Panel:} Simulated cosmic-ray trigger rate for several different atmospheric models. Here each model includes an increasing density of aerosols in the first 2 km of atmosphere. Cosinusoidal fits (of form $\mathrm{cos}^{n}(\theta)$) are made to each set of data as a function of zenith angle ($\theta$).   {\it Central Panel:} Real cosmic-ray trigger rate for the H.E.S.S. array versus zenith angle for 86 hours of data taken on the AGN PKS 2155-304 during August 2004. The fits for the the central panel are superimposed on the real data. {\it Right Hand Panel:} Zenith angle corrected real cosmic-ray trigger rate versus modified Julian date for  the AGN PKS 2155-304 taken during August 2004.}
\label{fig2}
\end{center}
\end{figure*}

\noindent Tables of optical depth versus Cherenkov photon wavelength and emission altitude were produced for a range of wind speeds from 0 m/s to 30 m/s.  These tables were then applied to a set of CORSIKA cosmic-ray simulations at various zenith angles between 0 and 60 degrees with a southern pointing, to match observations  taken on the AGN source PKS 2155-304, as shown in figure \ref{fig2}. To achieve a fit between data and simulation, an expected cosmic-ray trigger-rate for each atmosphere was derived for the H.E.S.S. array, based upon the spectra given in \cite{Wiebel} and using the sim\_telarray package \cite{KonradSim}. By matching the trigger-rate from simulations and real data, taking into account zenith angle dependence effects, an appropriate atmospheric model can be selected\footnotemark[1].\newline

\footnotetext[1]{Zenith angle corrections are derived from simulation and are to first order taken to be independent of atmospheric clarity.  }
\noindent As can be seen from the real cosmic-ray trigger rate data (centre and right panels of Figure  \ref{fig2}), it turns out that the  data can be separated into 3 classes, corresponding to MODTRAN4 atmospheric model wind speeds of 17.5, 20.0 and 22.5 m/s. 

\section{Higher Altitude Scattering Particles}

\noindent The ceilometer data suggest that the atmospheric conditions of the PKS 2155-304 data set under investigation can be completely described by normal weather conditions plus an additional low level aerosol population as described before. However, the ceilometer has a limited range ($\sim$7.5 km), and thus (additional) high-level aerosol layers or clouds can strictly speaking not be excluded by this instrument. Even worse, $\sim 27 \%$ of the data have no simultaneous ceilometer data at all; for these data, we solely have to rely on the classification by the cosmic ray trigger rate. We were therefore keen to develop an independent check for the eventual presence of high altitude absorbers, using the Cherenkov telescope data themselves. Such high-level absorbers might result in cosmic-ray trigger rates mimicking those resulting from low-level aerosols. But in this case, the correction applied to the gamma-ray data would likely be incorrect: Due to the different penetration depths of the two types of shower, absorbers in the altitude range of Cherenkov photon emission would likely affect gamma-ray and cosmic-ray air-showers differently (see also Figure \ref{3figs}).

\noindent The method that was developed is based on the comparison of the mean of the distribution for the reconstructed depths of shower maximum for cosmic-rays and gamma-rays, both real and simulated, under the application of different atmospheric models. The first step in this procedure was to compare the quality of fit for the reconstructed depth profiles under different atmospheric models. In Figure 6, a comparison of the reconstructed depth profile is shown for both observed and simulated gamma-rays at 20 degrees from zenith. Data from all zenith angles show that by utilising a Gaussian fit around shower maximum, the mean depth can be reliably compared to assess the similarity between simulations and real data. What is also clear is the fact that the mean of the distribution is not a measure of overall atmospheric clarity, but (as we shall shortly see) is correlated with the location of the aerosol population, allowing the telescope array to be used as a crude lidar system.

\begin{figure}[h]
\begin{center}
\includegraphics[clip=,angle=0,width=0.45\textwidth,height=0.25\textheight]{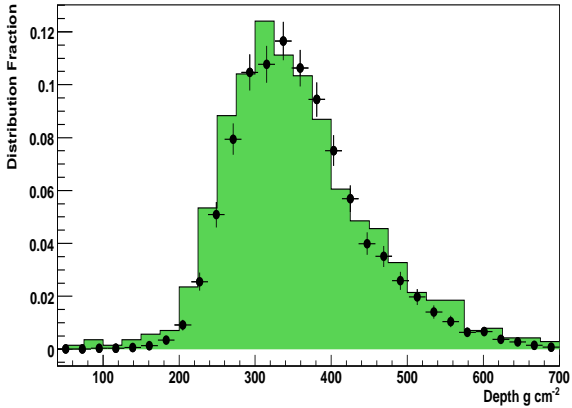}
\caption{Distributions of reconstructed depth of shower maximum for real background data (bar chart) and simulation (points) for data close to 20 degrees zenith angle for data picked by trigger rate to require the  22.5 m/s wind speed model. A Gaussian fit is sufficient to derive a measure for the mean depth of the distribution. }
\label{figcomp}
\end{center}
\end{figure}

\noindent To calibrate this procedure, a set of simulations with cloud layers at high altitudes was produced using MODTRAN4. These simulated atmospheres represent conditions which could in principle also have occurred during data-taking, as they result in cosmic-ray trigger rates that are similar to the low-level aerosol models. Each of the models includes a layer of thickness 1 km, the opacity of which is tuned to match the average observed cosmic-ray trigger rate. The base altitude is then varied for 6 different layers between 6 and 11 km above sea-level in 1 km steps. As shown in Figure \ref{figcomp}, by comparing the reconstructed shower depth for gamma-rays derived from real data and simulations, these high altitude models are considerably less favoured than the simple low-level aerosol models. Comparison of cosmic-ray shower depth simulations to real data also show agreement with the low-level aerosol model, whilst rejecting the high-level absorbers. 

\noindent High-level clouds such as cirrus clouds that could have well been present in some of the data are hence rejected in the entire data set. High-level aerosols are also not found, since their presence should have resulted in according shower height imprints, which are not seen. The non-detection of high-altitude aerosols does indeed not come as a big surprise. Tropospheric aerosols are typically limited to a few kilometers above ground (e.g. the Calima caused by Sahara dust is limited to $<$5.5 km \cite{Dorner}). Stratospheric aerosols at an altitude range of $\simeq$ 12km - 35km are normally only expected in case of volcanic activity \cite{stratosphere}.

\begin{figure*}[t]
\begin{center}
\includegraphics[width=0.95\textwidth,height=0.35\textheight,clip]{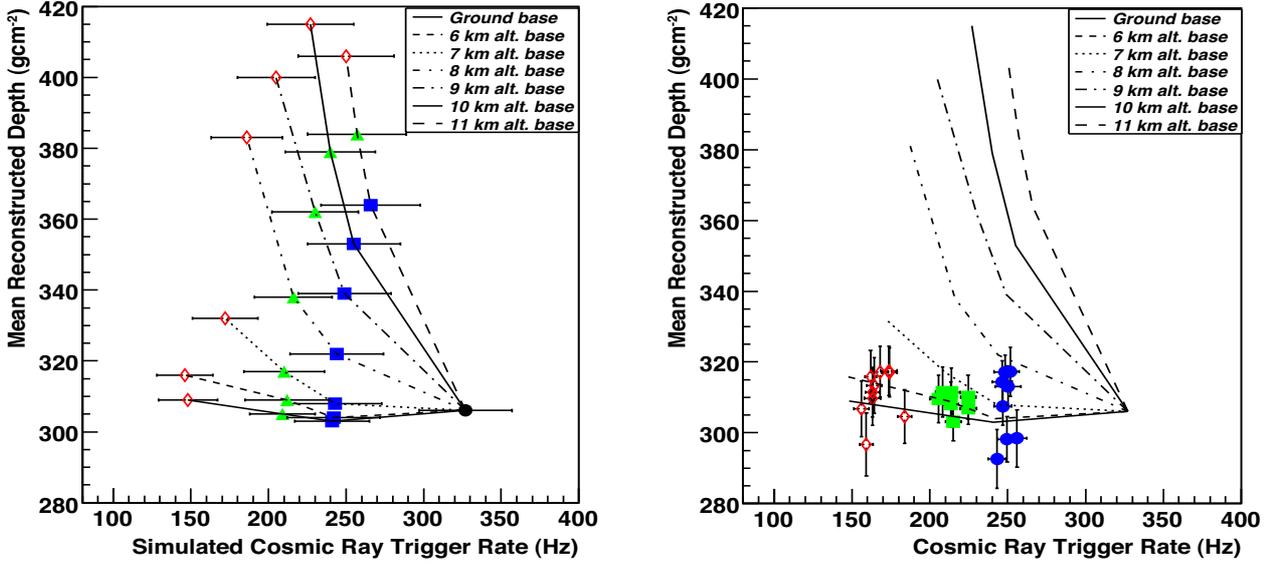}
\caption{\label {fig2a} The left panel shows the mean of reconstructed depth from a Gaussian fit for gamma-ray shower simulations at 20 degrees zenith-angle versus telescope trigger-rate. The lower points  show the results for the 10.0 (black), 17.5 (blue), 20.0 (green), and 22.5 (red) m/s wind speed models, with the other points showing the result for atmospheres with increasing altitude of the cloud layer, with lines connecting the same altitudes. These lines are reproduced on the right hand plot, which shows the  real mean reconstructed depth for gamma-ray data on PKS 2155-304 taken during 2004 at zenith angles between 15 and 25 degrees.  The data show no indication of high level clouds, thus independently confirming the ceilometer results.} 
\end{center}
\end{figure*}

\section{Gamma Ray Reconstruction}
\noindent This section discusses the effects on gamma-ray reconstruction due to the presence of low-level aerosols. Atmospheric models that include low-level aerosols were applied to a database of shower and telescope simulations, then lookup tables were derived using the method discussed in \cite{CrabPaper}.\newline

\noindent Lookup tables are generated by firstly reconstructing the core location of the gamma-ray shower by finding the intersection point for all the axes of the elliptical images seen in each telescope; secondly additional shower parameters are calculated telescope-wise and then averaged over all telescopes participating in the event. For example, to reconstruct the energy of events (either simulated or measured) one would firstly calculate additional parameters such as the distance of the shower core from the telescopes and the brightness of the image seen by the telescope cameras. Then the reconstructed energy (and an associated error) can be read from a lookup table derived from simulations.\newline

\noindent The key parameters for gamma-ray reconstruction are the effective area, the reconstructed energy, and the reduced mean-scaled length and width (MSL and MSW) \cite{CrabPaper}. How low-level aerosol density affects each of these variables is discussed below. 

\subsection{Effective Area}
\noindent When considering the changing levels of aerosol density, the naive guess would be to assume that mostly dimmer, less energetic events are affected. However, in reality the dimming effect has two consequences. Firstly, when dust levels are increased, more Cherenkov light is absorbed  which results in events close to the threshold of the system failing to trigger the system, but also in lowering the trigger yield across the entire energy range (Effect A). Secondly, the reconstructed energies of all events (regardless of their energy) will be biased towards low energies (Effect B) if this effect is not corrected for in the analysis.\newline

\begin{figure*}[t]
\begin{center}
\includegraphics[width=0.90\textwidth,height=0.34\textheight,clip]{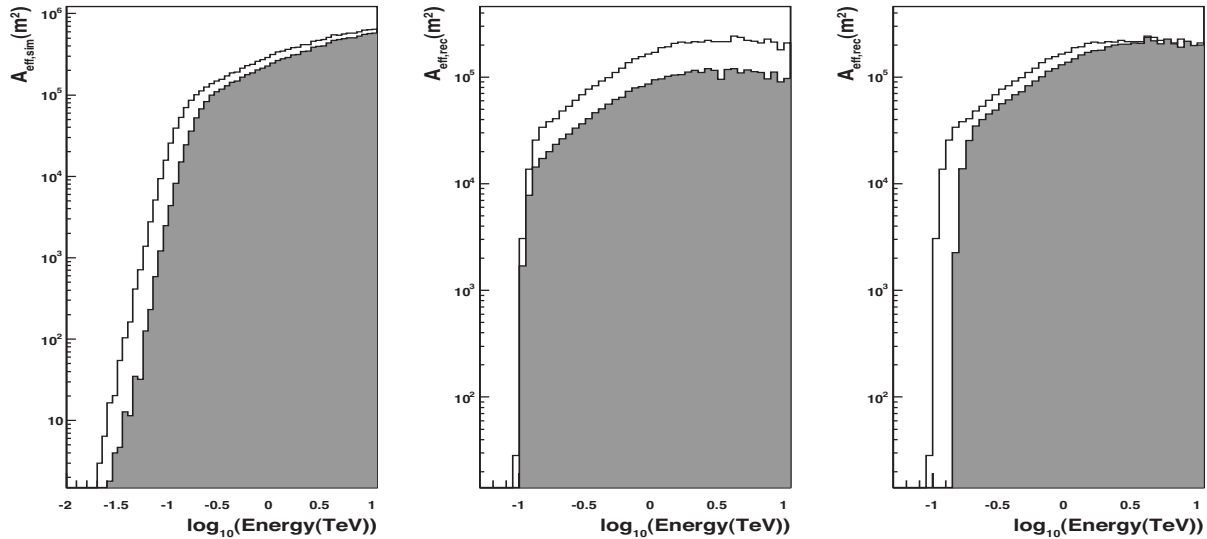}
\caption{\label {ea2}  {\it Left Panel:} The triggering effective area (as a function of simulated energy) calculated using an atmospheric model with a wind speed of 22.5 metres per second (shaded) and that calculated using a standard atmospheric model with wind speed 10 metres per second (non-shaded), at 20 degrees from zenith. {\it Central Panel:} The effective area is shown as as a function of reconstructed energy, for events passing standard background rejection shape cuts and lying within 1 degree of the source position. Here the standard energy lookup table is used for the both the standard atmospheric model simulations and the 22.5 metres per second model simulations. {\it Right Panel:} The same data are shown as the central panel, here different energy lookups (based on the appropriate simulations) are used for the both the standard atmospheric model simulations and the 22.5 metres per second model simulations, respectively.}
\end{center}
\end{figure*}

\noindent When illustrating both effects on the effective area, one should recall that in TeV Cherenkov astronomy, the effective area comes in two flavours, as a function of the simulated energy (A$_{\mathrm{eff,sim}}$(E)), and as a function of the reconstructed energy (A$_{\mathrm{eff,rec}}$(E)). Both functions are derived from simulations and are defined as

\begin{eqnarray}
A_{\mathrm{eff,sim}}(E)=\frac{N_{\mathrm{trig}}(E)}{N_{\mathrm{sim}}(E)}\pi r^{2},\\
A_{\mathrm{eff,rec}}(E)=\frac{N_{\mathrm{pass}}(E')}{N_{\mathrm{sim}}(E)}\pi r^{2}.
\end{eqnarray}

\noindent N$_{\mathrm{sim}}$ is the number of simulated events of energy E. The location of the simulated shower cores is randomly placed within a circle of radius r, chosen to be large enough so that showers at the border of the circle do not trigger the telescope array. N$_{\mathrm{trig}}$ is  the number of events which trigger the array system, and  N$_{\mathrm{pass}}$ stands for the number of events which in addition survive selection cuts to reject background events. \newline

\noindent To illustrate the impact of effect A, the left hand panel of Figure 7 shows the triggering effective area as a function of simulated energy, A$_{\mathrm{eff,sim}}$(E), calculated for an atmospheric model with a wind speed of 22.5 m/s (shaded), in comparison to that for a standard atmospheric model at a standard lower wind speed of 10 m/s (clear). The fact that dim images fail to trigger the instrument is most clearly seen at low energies. However, since higher energy events with large core distances to the telescope system contain dim images as well, the effective area is decreased across the entire energy range.\newline

\noindent To actually reconstruct source spectra from measured spectra, the finite energy resolution of the instrument has to be taken into account. This is done by replacing the number of triggered events with {\em simulated} energy E in A$_{\mathrm{eff,sim}}$(E) (equation 1) by the number of triggered events with {\em reconstructed} energy $E'$ in A$_{\mathrm{eff,rec}}$(E) (equation 2). As explained above, lookup tables give the reconstructed shower energy as a function of the distance of the shower core from the telescopes and the brightness of the Cherenkov images.\newline

\noindent The impact of effect B is demonstrated in the central panel of Fig.\,\ref{ea2}, where the effective area is shown as a function of reconstructed energy, A$_{\mathrm{eff,rec}}$(E), here for events which pass shape and distance cuts. Equation 2 implies that the effective area depends on the assumed source energy spectrum used to calculate A$_{\mathrm{eff,rec}}$(E). In regular analyses, this has only a mild impact, provided that the simulated photon spectrum is not too different from the source spectrum.\footnote{Normally this means that the photon index $\Gamma$ of the simulated power-law spectrum $I(E) \propto E^{-\Gamma}$ is similar to the source spectrum ($\Delta \Gamma \lesssim 0.5$), and to reach high accuracy, the effective area may be recomputed after a first reconstruction.} In addition, the impact of the energy bias seen at the system energy threshold is suppressed by starting spectral analysis only above an analysis threshold where the energy bias is below, say, 10\%.\newline 

\begin{figure}[h]
\begin{center}
\includegraphics[width=0.40\textwidth,clip]{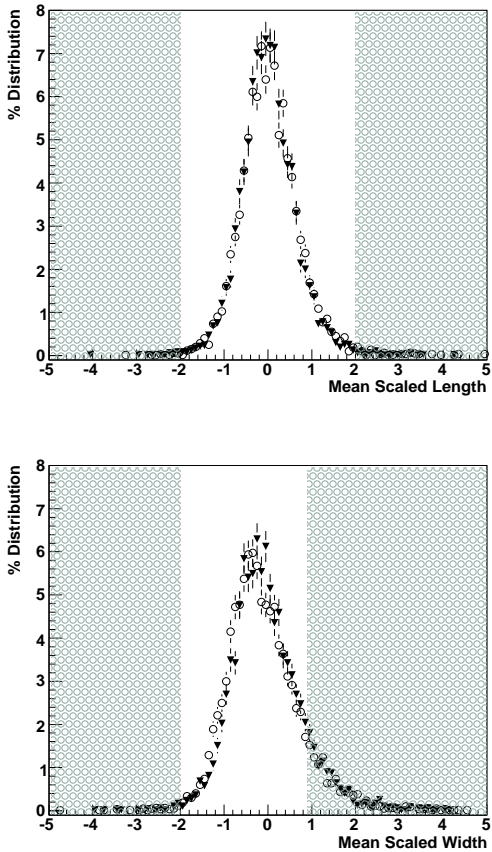}
\caption{\label {eff} Figure showing the MSL and MSW distributions for simulations at zenith for gamma-rays using an atmospheric model with wind speed parameters of 17.5 m/s (circles ) and 22.5 m/s (inverted triangles). The positions of the standard cuts (as detailed in \cite{CrabPaper}) are highlighted by the shaded areas. When selecting gamma-rays, values in these regions are rejected. }
\end{center}
\end{figure}

\noindent In the case at hand, however, an energy bias is seen across the entire energy range (Effect B, see also Fig.\,\ref{enres}). To illustrate this effect, the 22.5 m/s wind speed model effective area (shaded) was computed with an energy reconstruction using a standard atmosphere energy lookup table. This is compared to the standard, 10 m/s wind speed atmospheric model (clear). The use of the wrong energy reconstruction for the 22.5 m/s effective area reflects the normal, uncorrected approach to reconstruct data affected by low atmospheric transmission. Since the apparent energy thresholds of both effective areas are equal, a simple scaling factor corresponding to the ratio of the two effective areas seems at first glance appropriate to rescale reconstructed source fluxes, as discussed in the introduction. Such a scaling factor may be derived from the cosmic ray trigger rate. However, such a scaling factor depends on the actual source spectrum. Moreover, the scaling cannot account for the shape difference of the two effective areas, thus leading to a slight bending of the reconstructed spectra over the whole energy range, visible e.g. through a systematic shift in the reconstructed photon index.\newline

\noindent The right hand panel shows the same data as the central panel, but now the correct lookup tables are used also for the 22.5 m/s effective area. The actual shift in the energy threshold of the system under low-level aerosol conditions is recovered, and correct spectral shapes can be reconstructed using the appropriate effective area.\newline   

\noindent To conclude the section on the effective areas, we remark that the flattening of A$_{\mathrm{eff,rec}}$(E) above 2-3 TeV compared to A$_{\mathrm{eff,sim}}$(E) is caused by two effects: To provide a decent event reconstruction, only events with a core distance of 200 m  are used, which is a standard procedure. But in addition, only events up to 10 TeV were simulated in the presented study, leading to an artificial lack of reconstructed events near 10 TeV, because of the finite energy resolution. This does not affect any conclusion drawn in this paper, because spectra are only evaluated below 1 TeV (see Fig.\,\ref{spec}), and in addition the focus of this paper is on the comparison of spectra reconstructed with different effective areas.

\subsection{Energy Resolution}
\begin{figure*}[tb]
\begin{center}
\includegraphics[width=0.90\textwidth,height=0.3\textheight,clip]{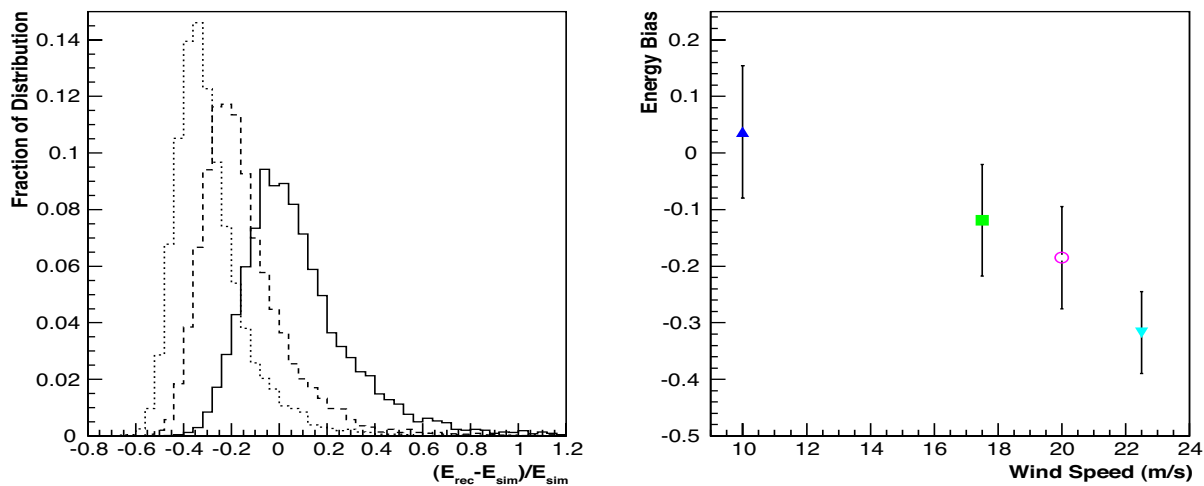}
\caption{\label {enres} The effect of increasing the low-level aerosol density on the energy resolution of the H.E.S.S. system. {\it Left Hand Panel:} The distribution of fractional energy offset for a set of simulated gamma-ray showers subject to both the standard (10 m/s) atmospheric model (solid line) and models with different wind-speed parameters (i.e. increased low-level aerosols density) 20.0 (dashed line) and 22.5 (dotted line) m/s. The lookup tables used in the derivation are those based on a standard atmospheric model only. {\it Right Hand Panel:} The resultant mean value of  ($E_{\mathrm{rec}}-E_{\mathrm{sim}})/E_{\mathrm{sim}}$ as derived from a Gaussian fit to the data  shown in the left hand panel versus the wind speed model used (Key: 10 m/s (triangle); 17.5 m/s atmosphere (square); 20.0 m/s atmosphere (circle); 22.5 m/s atmosphere (inverted triangle)). }
\end{center}
\end{figure*}

\noindent Effect B discussed in the previous section is the result of the dimming of all images recorded by the telescope cameras, when the lower level aerosol density is increased in simulations. Thus when estimating the energy from simulated lookup tables, if the effect of a changed atmosphere is not accounted for, too low an energy will be reconstructed. This creates a systematic bias in the energy reconstruction, as shown in Figure \ref{enres}.  Here the fractional energy offset is defined as the fractional difference between simulated energy and that reconstructed from lookup tables, namely ($E_{\mathrm{rec}}-E_{\mathrm{sim}})/E_{\mathrm{sim}}$.  After accounting for the atmospheric effects (i.e. using the appropriate reconstruction for the given atmospheric simulation), the bias is reduced to zero, and the resolution (derived as the standard deviation of a Gaussian fit to the fractional energy offset) is close to the energy resolution for a standard atmosphere. 

\subsection{Cut Efficiency} 
\noindent We now discuss the effect of increased low-level aerosol density on the efficiency of gamma-ray retention after cuts to remove background, under a standard H.E.S.S. analysis \cite{CrabPaper}. Simulation studies for a point source indicate no significant correlation between low-level aerosol density and cut efficiency; the shape parameters remove $>$95$\%$ of the background, whilst retaining $>60\%$ of the gamma-ray signal irrespective of the low-level aerosol densities present. Figure \ref{eff} illustrates this point. Here a lookup table was produced from a standard (10 m/s) set of simulations in order to derive the distribution of MSL and MSW for gamma-ray simulations at zenith using atmospheric models with a wind speed parameter of 17.5 and 22.5 m/s. No change with atmospheric model is apparent; this insensitivity is largely due to the fact that the MSL and MSW parameters are scaled to the image intensity.  Although there is no apparent dependence on low-level aerosol density, MSL and MSW lookup tables which account for changes in low-level aerosol density are used in our later derivation of fluxes.\newline

\noindent The previous discussion implies that the dominant effect of low-level aerosols besides the obvious flux reduction ‚is to bias energy reconstruction due to the lowering of the Cherenkov yield at the telescope. In typical power-law type spectra, the resulting excess of reconstructed events at low energies leads just to a very moderate softening of spectra if unaccounted for (values will be discussed below). Conveniently, further effects from shape cuts or energy resolution are marginal.

\section{Reconstructing the Flux of PKS 2155-304}
\noindent By applying the analysis tuned to different atmospheric models we can now attempt to resurrect H.E.S.S. data from a multi-wavelength campaign on the AGN PKS2155-304 in order to accurately reconstruct the spectral energy distributions. A large fraction of the data failed standard run selection criteria due to the effect of increased low-level aerosol density. These data were taken in the context of a multi-wavelength campaign which involved contemporaneous observations with 7 other telescopes or observatories covering a wavelength range from infra-red to X-rays. Therefore it was important to resurrect and correctly calibrate the H.E.S.S. data affected by low-level aerosols.\newline

\noindent  Over the 22 nights on which data were taken,  three distinct classes of low telescope trigger rate were observed,  as shown in Figure \ref{fig2}. These data could therefore be subdivided by ceilometer signal and cosmic-ray rate into 3 distinct classes, where either the 17.5 m/s, 20.0 m/s or 22.5 m/s wind speed model was appropriate. Using the methods outlined above, the low-level aerosol population was verified to be the dominant atmospheric contaminant. Table \ref{tab3} shows that the most frequently applied atmospheric model was that with a wind speed parameter of 22.5 m/s, accounting for 50$\%$ of the total exposure. The other two models were applied to the remaining 50$\%$ of the data-set.\newline

\begin{table}[tbh]
\begin{center}
{\footnotesize
\begin{tabular}{| c | c | c |}
\hline
Wind Speed&Exposure (Hours)& $\%$ of \\
& &Total Exposure\\
\hline
17.5&23.7&27.4\\
20.0&19.5&22.6\\
22.5&43.4&50.0\\
\hline
\end{tabular}
}
\end{center}
\caption{Table showing the exposures of the AGN PKS 2155-304 as a function of atmospheric conditions during August 2004.}   
\label{tab3}
\end{table}

\begin{figure*}[h]
\begin{center}
\includegraphics[width=0.95\textwidth,height=0.3\textheight,clip]{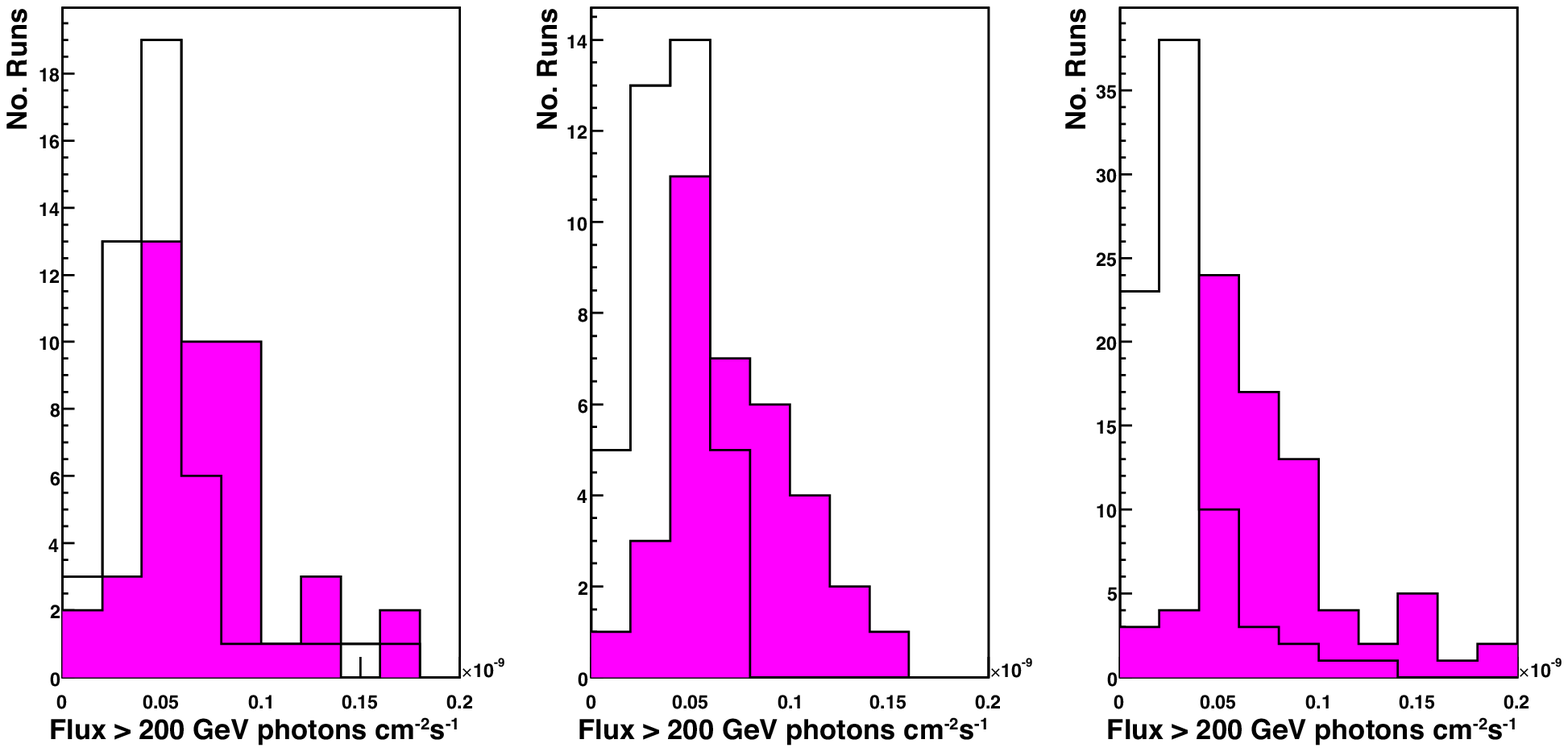}
\caption{\label {flux}The  distribution of integral flux for the AGN PKS  2155-304 above 200 GeV derived from 28 minute runs is plotted before (open histograms) and 
after (filled histograms) the application of corrections for low-level dust.  {\it Left Panel:} Atmospheric model 17.5 m/s wind speed parameter.  {\it Centre Panel:} Atmospheric model 20.0 m/s wind speed parameter.  {\it Right Panel:} Atmospheric model 22.5 m/s wind speed parameter. }
\end{center}
\end{figure*}

\begin{figure*}[h]
\begin{center}
\includegraphics[width=0.95\textwidth,height=0.3\textheight,clip]{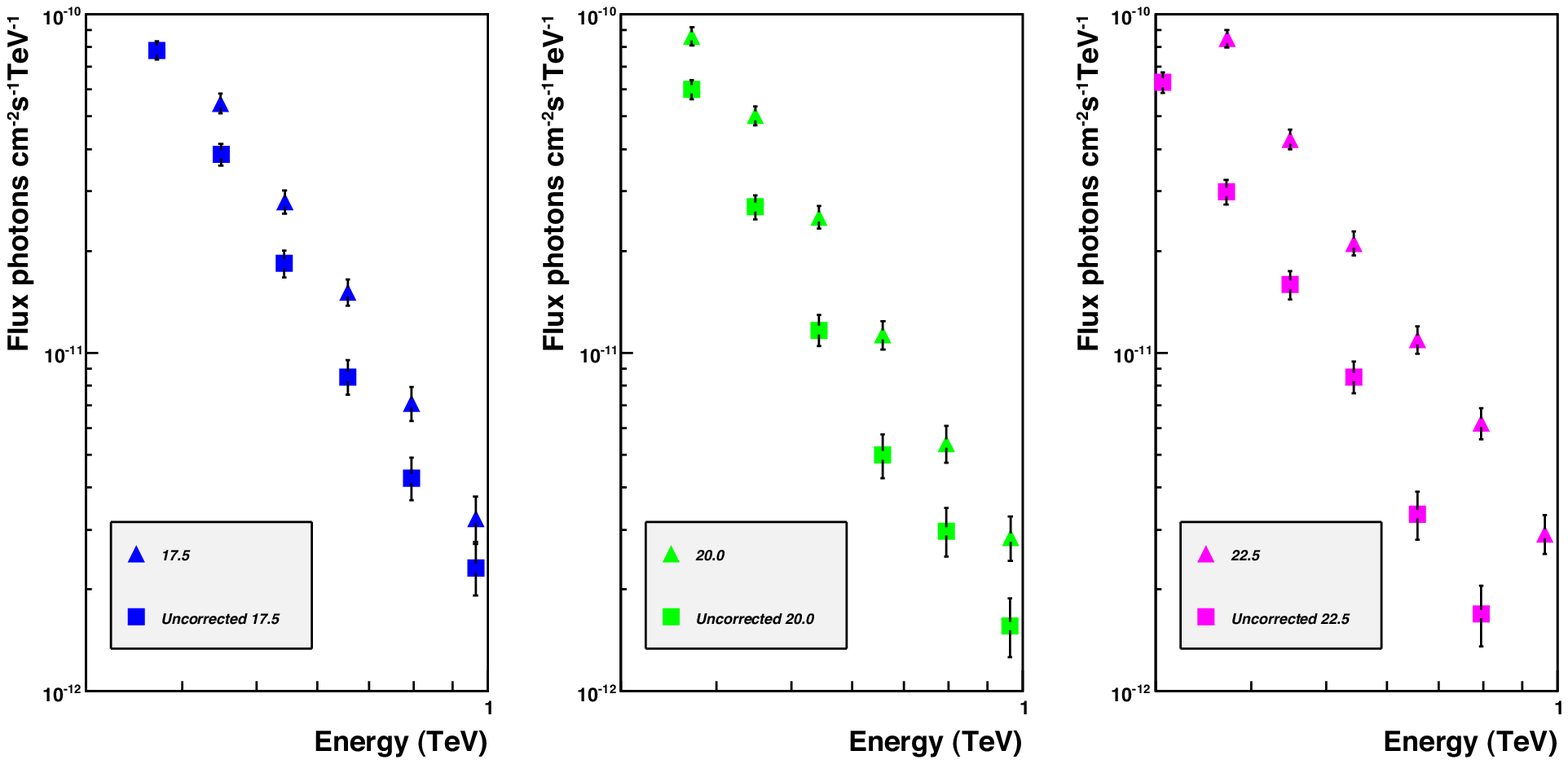}
\caption{\label {spec}The  uncorrected (squares) and corrected (triangles) differential spectra for the three atmospheric models  is shown between 300 GeV and 1 TeV.  Above 1 TeV statistical errors prevent a meaningful comparison. {\it Left Panel:} Atmospheric model 17.5 m/s wind speed parameter.  {\it Centre Panel:} Atmospheric model 20.0 m/s wind speed parameter.  {\it Right Panel:} Atmospheric model 22.5 m/s wind speed parameter. }
\end{center}
\end{figure*}

\noindent By creating a database of 18 million gamma-ray showers at each of eight zenith angles (0, 20, 30, 40, 45, 50, 55 and 60 degrees) and simulating the H.E.S.S. telescope system response to these showers, a set of lookup tables for energy, effective area, MSL and MSW (as discussed in \cite{CrabPaper}) was produced for each wind speed parameter (17.5, 20.0 and 22.5 m/s).  Fluxes and spectra were then derived for each subset of data using the appropriate lookup table as suggested by the cosmic-ray trigger rate. The results for flux and spectra are shown in Figures \ref{flux} and \ref{spec}, respectively.

 \section{Discussion}
\noindent  We are confident that the presented method yields the correct energy spectra for PKS 2155-304. Given however the variable nature of the source, neither flux nor photon index is expected to match results obtained from observations in other years, therefore the results cannot be verified by such comparisons. A constant gamma-ray source of sufficient brightness that could be used as calibrator was not observed with HESS during the aerosol contamination period. We were however able to identify a small set of Crab Nebula data that were taken under similar atmospheric conditions, despite the fact that Crab observing season is in the other half of the year. Indeed, those ~1.5 hours of observations can best be described -- according to cosmic-ray trigger rate and ceilometer return signal -- by the 17.5 m/s wind speed parameter atmospheric model. The parameters for a power-law fit to the differential spectrum of this Crab Nebula data are shown in Table \ref{tab4}, along with the fit parameters for the PKS 2155-304 dataset. The application of the appropriate reconstruction to the Crab data leads to a very significant increase in flux normalisation ($\Delta$ I$_0$ $\simeq $25$\%$), which after correction is in perfect agreement with the published values derived from unaffected data \cite{CrabPaper}. For the photon index $\Gamma$, unfortunately, a meaningful statement is not possible, because of the statistical error and the smallness of the correction ($\Delta \Gamma \simeq$ 0.03). Both uncorrected and corrected values for $\Gamma$ are within errors in agreement with the published value.

\noindent The differences in flux normalisation and photon index between uncorrected and corrected numbers for the PKS 2155-304 data is larger than those for the Crab data. The difference in flux normalisation for PKS 2155-304 shows the expected strong correlation with atmospheric absorption, $\Delta I_0 \sim$ 38$\%$ (17.5 m/s) - 64$\%$ (22.5 m/s), while $\Delta \Gamma \simeq$ 0.13 for all three subsets. In general, we believe that the different corrections for the Crab and PKS 2155-304 data sets are related to the different photon indices of those sources, since the corresponding reconstructed energy effective areas (shown in figure \ref{ea2})  when weighted by the different photon indices differ. 
\noindent Overall, it is fairly reassuring that the magnitude of the photon index correction is in the same range as the systematic error quoted for H.E.S.S. photon indices ($\Delta \Gamma \sim$ 0.1), which was estimated for data taken under normal good weather conditions.

\begin{table*}[b]
\begin{center}
{\footnotesize
\begin{tabular}{| c | c | c | c | c | c | c |}
\hline
Source & Type & Atm & Lookup & I$_{0}\times 10^{-12}$ & $\Gamma$ & $\chi^{2}$/NDF \\
 & Model & Used & & & & \\
\hline
Crab Nebula & Uncorrected & 17.5 & std & 26.0$\pm$1.0 & 2.62$\pm$0.07 & 0.67 \\
Crab Nebula & Corrected & 17.5 & 17.5 & 35.0$\pm$2.0 & 2.59$\pm$0.07 & 0.77 \\
\hline
\hline
PKS 2155-304 & Uncorrected & 17.5 & std & 2.1$\pm$0.2 & 3.50$\pm$0.07 & 2.3 \\
PKS 2155-304 & Corrected & 17.5 & 17.5 & 3.4$\pm$0.3 & 3.38$\pm$0.07 & 2.3 \\
\hline
PKS 2155-304 & Uncorrected & 20.0 & std & 1.4$\pm$0.1 & 3.63$\pm$0.05 & 1.8 \\
PKS 2155-304 & Corrected & 20.0 & 20.0 & 2.7$\pm$0.1 & 3.50$\pm$0.04 & 0.86 \\
\hline
PKS 2155-304 & Uncorrected & 22.5 & std & 0.9$\pm$0.1 & 3.60$\pm$0.05 & 1.52 \\
PKS 2155-304 & Corrected & 22.5 & 22.5 & 2.5$\pm$0.1 & 3.47$\pm$0.04 & 1.70 \\
\hline
\end{tabular}
}
\end{center}
\caption{Parameters are given for a power law fit to the differential energy spectrum  $I(E)=I_{0}E^{-\Gamma}$, where $I_{0}$ is measured in photons cm$^{-2}$ s$^{-1}$ TeV$^{-1}$.  }
\label{tab4}
\end{table*}

\noindent Although the technique presented in this paper appears to be a viable method for resurrecting data affected by increased low-level aerosol density, there are a few shortcomings. Firstly, the atmospheric model is selected based upon a comparison of cosmic-ray simulation and background data from the telescope. Given uncertainties in the cosmic-ray flux at 1 TeV, simulated fluxes have a systematic uncertainty of $\sim$10$\%$. An independant measurement of the atmospheric transmission profile would overcome this reliance on cosmic-ray data.  Secondly, due to the large number of shower and telescope simulations required, the method is highly computationally intensive, with over 16 CPU years spent on the data-set discussed in this paper (67 CPU days per hour of exposure). Thirdly, given the seasonal dependence of the low-level aerosol densities, there are only parts of the year when sources visible at the H.E.S.S. site are affected by this form of low-level haze. This method is therefore not routinely applied to H.E.S.S. data. Instead any data affected by measured adverse atmospheric conditions is removed from the analysis.\newline

\noindent A clear step forward when tackling the problem of low-level aerosol populations is to place lidars at the telescope sites from which atmospheric transmission can be extracted directly. For this purpose two high power lidars were installed in (2007/2008) at the H.E.S.S. site. These systems allow comparisons between simulations of cosmic rays, gamma-rays and real data, which will be used as an independent cross check of the inferred atmospheric model. Cross comparisons between the two systems will serve as a further check of atmospheric assumptions. 

\section{Conclusion}
\noindent A detailed discussion of the effects of low-level aerosol density on the H.E.S.S. system has been presented, and a method to correct for these effects in energy and flux has been demonstrated to be successful on observational data of the AGN PKS 2155-304. Without correction, if data with significant low-level aerosol densities are used, substantially too low fluxes are reconstructed, and a systematic softening of the reconstructed spectra is observed. This work, for the first time, combines direct measurements of the profile of low-level atmospheric aerosols with simulation and accounts for atmospheric aerosol population during flux derivation, allowing the resurrection of nearly 100 hours of otherwise unusable data, a significant fraction of the observation season for 2004. A follow up paper combining corrected H.E.S.S. data with contemporaneous multi-wavelength observations of PKS 2155-304 during 2004 is currently in preparation. Two new lidar instruments recently placed on the H.E.S.S. site will further improve our knowledge of the atmosphere and serve to both reduce systematic errors on measured flux and increase the active lifetime of the experiment. 

\vspace{4ex}
\par{\noindent {\it Acknowledgements}  The authors would like to acknowledge the support of their host institutions, and in addition we are grateful to Werner Hofmann, Michael Punch, Paula Chadwick \& Michael Daniel for their helpful comments. S.N. and C.R. acknowledge support from the Science and Technology Facilities Council of the UK. G.P. acknowledges support by the German Ministry for Education and Research (BMBF) through DESY grant 05CH5VH1/0 and DLR grant 50OR0502.}

\vspace{2ex}

\end{document}